\documentclass[aps,prb,twocolumn,showpacs,superscriptaddress,amsmath,amssymb,floatfix,10pt]{revtex4-1}  % for review and submission
\usepackage{graphicx}  % needed for figures
\usepackage{dcolumn}   % needed for some tables
\usepackage{bm}        % for math
\usepackage{amssymb}   % for math
\usepackage{framed}
\usepackage{amsmath}
\usepackage{multirow}
\usepackage{hhline}
\usepackage{color}
\definecolor{bluegreen}{rgb}{0,0.2,0.8}
\usepackage{subfigure,amsmath,verbatim,moreverb}
\usepackage{tabularx}
\usepackage{adjustbox}
\usepackage{lipsum}
\usepackage{longtable}

\begin{document}

\widetext
\title{Efficient lattice constants and energy band gaps for condensed systems from a meta-GGA
level screened range separated hybrid functional}

\author{Subrata Jana}
\email{subrata.jana@niser.ac.in}
\affiliation{School of Physical Sciences, National Institute of Science Education and Research, HBNI, 
Bhubaneswar 752050, India}
%\author{Bikash Patra}
%\affiliation{School of Physical Sciences, National Institute of Science Education and Research, HBNI, 
%Bhubaneswar 752050, India}
\author{Abhilash Patra}
\affiliation{School of Physical Sciences, National Institute of Science Education and Research, HBNI, 
Bhubaneswar 752050, India}
%\author{Rabeet Singh}
%\affiliation{School of Physical Sciences, National Institute of Science %Education and Research, HBNI, 
%Bhubaneswar 752050, India}
\author{Prasanjit Samal}
%\email{psamal@niser.ac.in}
\affiliation{School of Physical Sciences, National Institute of Science Education and Research, HBNI, 
Bhubaneswar 752050, India}

\date{\today}

\begin{abstract}

A meta generalized gradient level screened range-separated hybrid functional is developed for
solid-state electronic structure theory. Assessment of the present range-separated hybrid
functional for solid-state lattice constants and band gaps indicate that the present
functional can be used for describing those properties efficiently in meta-GGA level.
Specifically, the performance of the present functional for band gap of solids 
indicates that the present meta-GGA level screened hybrids functional is quite productive
beyond the GGA level. The most appealing feature of the present formalism is that a method has
been suggested which is based upon an accurate semilocal functional. 

\end{abstract}

\maketitle

\section{Introduction}

The Kohn-Sham (KS) formalism of density functional theory (DFT)~\cite{ks65,engelbook} 
is one of the most widely used and accurate theoretical framework for electronic structure
calculations of condensed systems. In the KS formalism the system is effectively one electron
like where all the many electron interactions are taken care by exchange-correlation (XC)
energy or potential. The effective one electron potential in DFT, also best known as KS
potential is sum of the classical coulomb or Hartree potential ($v_H$), the
exchange-correlation potential ($v_{xc}$), and the external potential generated by the nuclei
(or ion) ($v_{ext}$): 
\begin{equation}
 v_{_{\mathrm{KS}}}(\mathbf{r}) =  v_H(\mathbf{r}) + v_{xc}(\mathbf{r})  +
 v_{ext}(\mathbf{r}).
\label{eq1}
\end{equation}

In Eq.(\ref{eq1}) the only unknown quantity is the $v_{xc}$, which need to be treated
approximately in DFT. Various approximations~\cite{jacob,lda,PW86,B88,LYP88, PW91,
B3PW91,PBE96,AE05,ZWu06,PBEsol,con4,con1,con3,con5,BR89,VSXC98,HCTH,PBE0,MO6L,TPSS03,revTPSS,
con2,SCAN15,Tao-Mo16} are proposed for last couple of decades to treat accurately the
$v_{xc}$. All these approximations are recognized through the Jacob's ladder~\cite{jacob},
where each rung of the ladder add an extra ingredients starting from local density
approximations~\cite{lda}. The generalized gradient approximations
(GGA)~\cite{PW86,B88,LYP88,PW91,PBE96,AE05,ZWu06,PBEsol,con4,
con1,con3,con5} and meta generalized gradient approximations
(meta-GGA)~\cite{BR89,VSXC98,HCTH,PBE0,MO6L,TPSS03,revTPSS,con2,SCAN15,Tao-Mo16} are next two
higher rung after LDA. The LDA, GGA and meta-GGA are widely used for performing electronic
structure calculations~\cite{PHao13,tmpccp,phmk,PBlaha09,FTran16,YMo16,YMoCPL,mapkw,smrkp,
mgga-vasp,jana} in DFT community for their semilocal nature. though the semilocal
approximations enjoy early success but due to lack of ``many electron self interaction
(MESI)'' and ``non-locality''~\cite{Yangreview} there are cases in which the performance of
semilocal approximations is not satisfactory. The non-locality within the 
density functionals approximations (DFAs) are introduced through the mixing of Hartree-Fock
(HF) exact-exchange either globally (global hybrids)~\cite{B3PW91,B3LYP,VNS03} or in range
separated scheme (range separated hybrids)~\cite{HSE03,HSE06,camb3lyp,lcwpbe,tpssrs,
ldars,ityh01,bikash,sjanapccp}. Though the global hybrid functionals are very popular in
quantum chemistry~\cite{B3PW91,B3LYP,VNS03} but they are not so popular for condensed matter
electronic structure theory~\cite{b3lypsolid}. Beyond the global and long range corrected
hybrid functionals, the range separated hybrids proposed using short range HF with DFAs are
very popular due to their very improved performance for the solid state systems, especially
in band gaps~\cite{HSE03,HSE06,mpsk,jmhk,shk,bgref,ottpk,sgsbk07,jvruk10,pmk08,kmktb10,sk08,
fppmk07,wvk08,stpkh07,khk09,hhk09,gamk07,hs04,hpsm05,hjs08,phsm06,binbs08,tranbg,galli1,
galli2,galli3,galli4}.

Designing a range separated hybrid functionals requires the exchange hole. The exchange hole
is proposed using Taylor series approximations~\cite{BR89} or density matrix
expansion~\cite{Tao-Mo16} or reversed engineered technique~\cite{tpssrs,tpsshole,cfs,cpp09}.
The popular Heyd-Scuseria-Ernzerhof (HSE)~\cite{HSE03,HSE06} functional is designed using 
the reversed engineered exchange hole of PBE functional. Beyond the GGA level screened range
separated hybrid functionals, the meta-GGA level screened range separated hybrid functionals
is also proposed recently by Tao et.al.~\cite{tpssrs} using by utilizing the TPSS exchange
energy functionals. Beyond the TPSS exchange energy funcional, very recently, 
Tao-Mo proposed an accurate semilocal functional~\cite{Tao-Mo16} for quantum chemistry and
solid state system using density matrix expansion based semilocal exchange hole with the
slowly varying fourth order gradient approximation. Motivated by the TM functional and its
underlying construction we propose an screened range separated hybrid functional to be used
for condensed matter systems. In designing the present screened range separated hybrid
functional we utilize the local density approximation based exchange hole for the short range
semilocal part with the short range HF. This is the possible conventional way to utilize the
TM functional in screened range separated hybrid functional
scheme for solid state systems bypassing its reversed engineered exchange hole. Though, very
recently, another way of inclusion of exchange hole in short range semilocal functional is
proposed~\cite{sjanapccp}, but, that scheme not perform satisfactory way as it is found in the
present work. Surprisingly, the only LDA exchange hole on the top of the TM
exchange-correlation functional performs efficiently in describing both
the lattice constants and band gaps of solids. In this  paper we design a screened range
separated hybrid functional in meta-GGA level and the performance of the present functional is
carried out for solid state lattice constants and band gaps using the projector-augmented-wave
method~\cite{paw1,paw2,vasp1,vasp2,vasp3,uspp} with the plane wave basis set. 

The present paper is organized as follows: In the following we will discuss about the
generalized KS potential to be used in the hybrid functional calculations. Following this we
will give the formulation of the present range separated
hybrid functional using semilocal exchange functional and short range HF. Next we will briefly
discuss the implementation of the developed range separated functional and its performance for
solid state lattice constants and band gaps.

\begin{table*}%[h]
\label{latt-cons}
\caption{Equilibrium lattice constant $a_0$ (in \AA)~ of
different solid structures using HSE06, SRSH-TM-TPSSc and SRSH-TM. All the experimental
reference values are
collected from ref.~\cite{bgref,YMo16}. The structures we consider here are A1 $=$
face-centered cubic, A2 $=$ diamond, A3 $=$ body-centered cubic, B3 $=$ zinc blende, and
B1 $=$ rock salt. The relative deviation (in percentage) of the individual species using each
functional is also given.}
\begin{tabular}{c  c  c  c  c  c   c   c  c c}
\hline\hline
Solids&HSE06&\%&SRSH-TM-TPSSc&\%&SRSH-TM&\%&Expt.\\
\hline 
C (A2)&3.548&-0.53&3.550&-0.48&3.545&-0.62& 3.567 \\
Si (A2)&5.432&0.04&5.420&-0.18&5.408&-0.40& 5.430 \\
Ge (A2)&5.676&0.42&5.653&0.02&5.636&-0.28&5.652 \\
SiC (B3)&4.346&-0.27&4.338&-0.46&4.332&-0.59&4.358 \\
BN (B3)&3.597&-0.28&3.603&-0.11&3.597&-0.28&3.607 \\
BP (B3)&4.519&-0.42&4.52&-0.40&4.509&-0.64& 4.538 \\
BAs (B3)&4.770&-0.15&4.766&-0.23&4.754&-0.48&4.777 \\
BSb (B3)&5.216&n/a&5.202&n/a&5.188&n/a&n/a \\
%AlN& & & && & &4.400 & & \\
AlP (B3)&5.470&0.18&5.461&0..02&5.448&-0.22&5.460 \\
AlAs (B3)&5.676&0.32&5.659&0.02&5.646&-0.21&5.658 \\
AlSb (B3)&6.151&0.24&6.130&-0.10&6.114&-0.36&6.136 \\
%GaN& & & & && & & & \\
$\beta-$GaN (B3)&4.521&-0.22&4.526&-0.11&4.516&-0.33&4.531 \\
GaP (B3)&5.464&0.29&5.463&0.27&5.446&-0.03&5.448 \\
GaAs (B3)&5.667&0.34&5.653&0.09&5.635&-0.23&5.648 \\
GaSb (B3)&6.099&0.05&6.075&-0.34&6.055&-0.67&6.096 \\
%InN& & && & & & & & \\
InP (B3)&5.921&0.94&5.923&0.68&5.903&0.63&5.866 \\
InAs (B3)&6.108&0.89&6.095&0.68&6.075&0.34&6.054 \\
InSb (B3)&6.516&0.57&6.496&0.26&6.473&-0.09&6.479 \\
ZnS (B3)&5.419&0.18&5.436&0.50&5.412&0.05&5.409 \\
ZnSe (B3)&5.693&0.44&5.699&0.55&5.676&0.14&5.668 \\
ZnTe (B3)&6.135&0.75&6.129&0.65&6.099&0.16&6.089 \\
CdS (B3)&5.880&1.06&5.924&1.82&5.893&1.29&5.818 \\
CdSe (B3)&6.133&1.34&6.164&1.85&6.133&1.34&6.052 \\
CdTe (B3)&6.543&0.97&6.568&1.36&6.533&0.81&6.480 \\
MgO (B1)&4.197&-0.24&4.195&-0.28&4.187&-0.47&4.207 \\
MgS (B3)&5.652&8.65&5.647&8.55&5.632&8.27&5.202 \\
MgSe (B1)&5.454&1.00&5.462&1.15&5.411&0.76&5.400 \\
MgTe (B3)&6.452&0.50&6.446&0.40&6.424&0.06&6.420 \\
CaS (B1)&5.698&0.16&5.722&0.58&5.699&0.17&5.689 \\
CaSe (B1)&5.938&0.37&5.966&0.84&5.939&0.38&5.916 \\
CaTe (B1)&6.369&0.33&6.404&0.88&6.369&0.33&6.348 \\
SrS (B1)&6.034&0.73&6.071&1.35&6.046&0.93&5.990 \\
SrSe (B1)&6.268&0.54&6.302&1.09&6.275&0.66&6.234 \\
SrTe (B1)&6.684&0.66&6.721&1.22&6.688&0.72&6.640 \\
BaS (B1)&6.432&0.67&6.487&1.53&6.454&1.02&6.389 \\
BaSe (B1)&6.656&0.92&6.707&1.70&6.673&1.18&6.595 \\
BaTe (B1)&7.057&0.71&7.115&1.54&7.075&0.97&7.007 \\
Ag (A1)&4.146&1.89&4.151&2.02&4.135&1.62&4.069 \\
Al (A1)&4.020&-0.30&3.980&-1.29&3.979&-1.31&4.032 \\
Cu (A1)&3.637&0.94&3.573&-0.83&3.573&-0.83& 3.603 \\
Pd (A1)&3.904&0.59&3.923&1.08&3.909&0.72&3.881 \\
K (A3)&5.32&1.82&5.297&1.38&5.273&0.92&5.225 \\
Li (A3)&3.466&-0.32&3.439&-1.09&3.440&-1.06&3.477 \\
LiCl (B1)&5.116&0.19&5.100&-0.12&5.076&-0.59&5.106 \\
LiF (B1)&4.015&0.12&3.973&-0.92&3.968&-1.05&4.010 \\
NaCl (B1)&5.613&0.32&5.556&-0.70&5.540&-0.98&5.595      \\
NaF (B1)&4.576&-0.71&4.513&-2.08&4.507&-2.21&4.609      \\
\hline\hline
%ME(\AA)&&&&&&&$-$\\
%MAE(\AA)&&&&&&&$-$\\
%\hline
%STDE(\AA)&&&&&&&$-$\\
%MRE(\%)&&&&&&&$-$\\
%MARE(\%)&&&&&&&$-$\\
%STDRE(\%)&&&&&&&$-$\\
%\hline\hline
\end{tabular} 
\end{table*}

\begin{figure*}%[h]
\begin{center}
\includegraphics[width=6.0in,height=3.0in,angle=0.0]{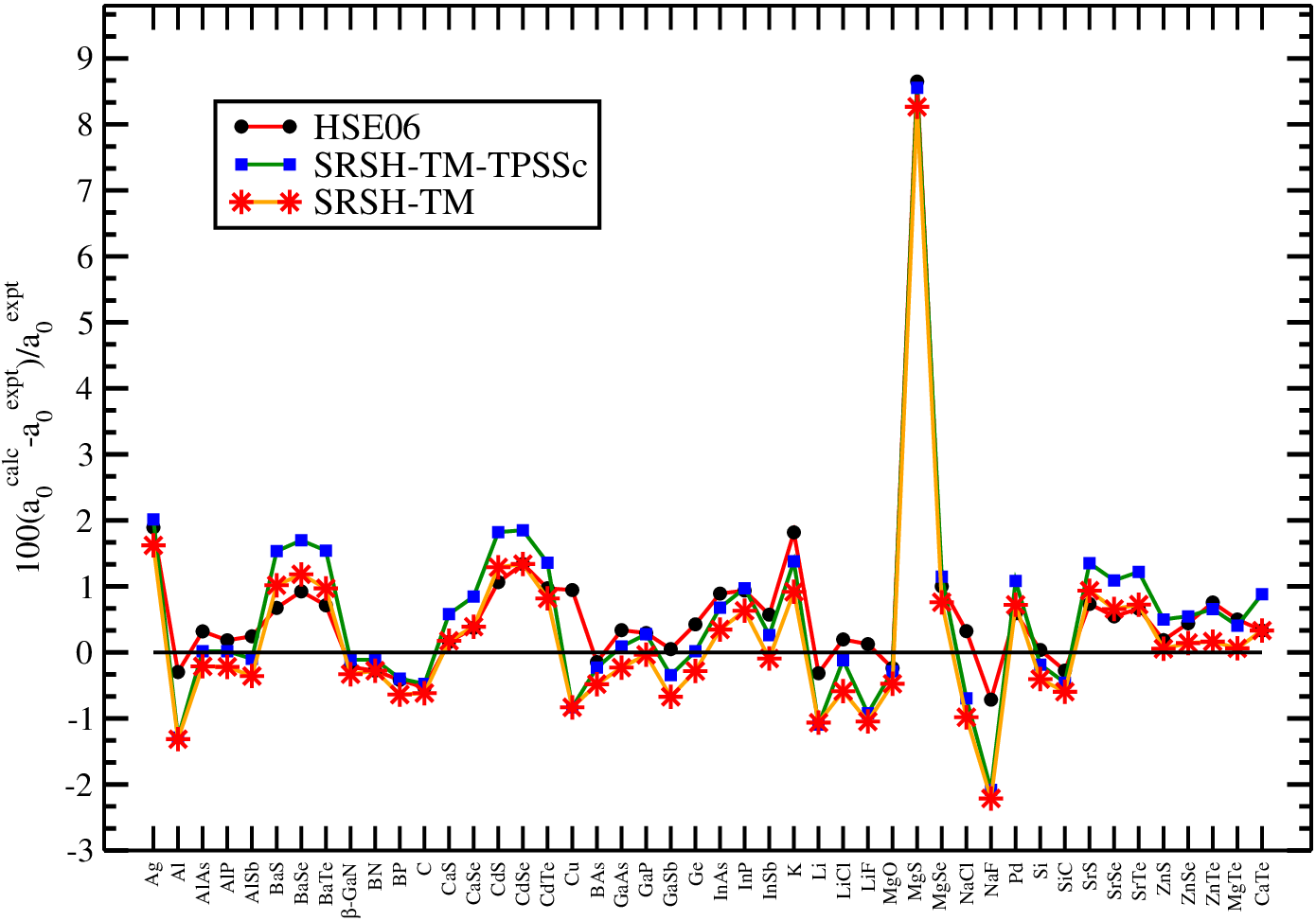} 
\end{center}
\caption{Relative deviation (in percentage) in the calculated
lattice constants with respect to the experimental (ZPAE-uncorrected) values (see Table I).}
\label{fig1}
\end{figure*}

\begin{table*}%[h]
\label{latt-cons}
\caption{Band gaps using different functionals calculated at the experimental lattice
constants. The structure considered here are A1 = rock salt, A2 = diamond, A3 = zinc blende.
Here, the experimental geometries and band gap values are 
taken from reference~\cite{tranbg}.
 }
\begin{tabular}{c  c  c  c  c  c   c   c  c c c c}
\hline%\hline
Solids&Space Group&Geometry(\AA)&HSE06&\%&SRSH-TM-TPSSc&\%&SRSH-TM&\%&Expt.\\
\hline
%Ne~(A1)     &Fm$\overline{3}$m	&4.470	&	& & & & & &21.48\\
%Ar~(A1)     &Fm$\overline{3}$m	&5.260	&	& & & & & &14.15\\
%Kr~(A1)	   &Fm$\overline{3}$m	&5.640	&	& & & & & &11.59\\
%Xe~(A1)	   &Fm$\overline{3}$m	&6.130	&	& & & & & &9.29\\
%SnTe~(A2)   &Fm$\overline{3}$m	&6.318	&	& & & & & &0.36\\
%LiH~(A2)	   &Fm$\overline{3}$m	&4.084	&	& & & & & &4.94\\
%LiF~(A2)	   &Fm$\overline{3}$m	&4.010	&	& & & & & &14.2\\
%LiCl~(A2)   &Fm$\overline{3}$m	&5.106	&	& & & & & &9.4\\
%NaF~(A2)	   &Fm$\overline{3}$m	&4.609	&	& & & & & &11.5\\
%NaCl~(A2)   &Fm$\overline{3}$m	&5.595	&	& & & & & &8.5\\
%KF~(A2)	   &Fm$\overline{3}$m	&5.347	&	& & & & & &10.9\\
%KCl~(A2)	   &Fm$\overline{3}$m	&6.293	&	& & & & & &8.5\\
MgO(A1)	   &Fm$\overline{3}$m	&4.207	&6.49	&-17.11 &6.84 &-12.64 &6.71 &-14.30 &7.83\\
%MgSe~(A1)   &Fm$\overline{3}$m	&5.400	&2.87	&16.29 & & & & &2.47\\
%CaO~(A2)	   &Fm$\overline{3}$m	&4.811	&	& & & & & &7.0\\
%CaF$_2$~(A2)   &Fm$\overline{3}$m	&5.463	&	& & & & & &11.8\\
BaS~(A1)	   &Fm$\overline{3}$m	&6.389	&3.06	&-21.13 &3.16 &-18.55 &3.09 &-20.36 &3.88\\
BaSe~(A1)   &Fm$\overline{3}$m	&6.595	&2.76	&-22.90 &2.91 &-18.99 &2.83 &-20.94 &3.58\\
BaTe~(A1)   &Fm$\overline{3}$m	&7.007	&2.27	&-26.29 &2.45 &-20.45 &2.38 &-22.72 &3.08\\
ScN~(A1)	   &Fm$\overline{3}$m	&4.500	&0.86	&-4.44 &1.07 &18.88 &1.02 &13.33 &0.9\\
AgCl~(A1)   &Fm$\overline{3}$m	&5.546	&2.43	&-25.23 &2.73 &-16.00 &2.66 &-18.15 &3.25\\
AgBr~(A1)   &Fm$\overline{3}$m	&5.772	&2.14	&-21.03 &2.59 &-4.42 &2.50 &-7.74 &2.71\\
C~(A2)	   &Fd$\overline{3}$m	&3.567	&5.29	&-3.81 &5.45 &-0.90 &5.36 &-2.54 &5.5\\
Si~(A2)	   &Fd$\overline{3}$m	&5.430	&1.17	&0.00 &1.41 &20.51 &1.30 &11.11 &1.17\\
Ge~(A2)	   &Fd$\overline{3}$m	&5.430	&0.82	&10.81 &1.04 &40.54 &1.01 &36.48 &0.74\\
%SiO$_2$~(A3)   &Fd$\overline{3}$m	&7.127	&	& & & & & &8.9\\
SiC~(A3)	   &F$\overline{4}$3m	&4.358	&2.35	&-2.89 &2.53 &4.54 &2.44 &0.82 &2.42\\
BN~(A3)	   &F$\overline{4}$3m	&3.616	&5.90	&-7.23 &6.19 &-2.67 &6.07 &-4.55 &6.36\\
BP~(A3)	   &F$\overline{4}$3m	&4.538	&2.01	&-4.76 &2.19 &4.28 &2.10 &0.00 &2.1\\
BAs~(A3)    &F$\overline{4}$3m	&4.777	&1.87	&28.08 &1.98 &35.61 &1.91 &30.82 &1.46\\
AlN~(A3)	   &F$\overline{4}$3m	&4.342	&4.72	&-3.67 &4.95 &1.02 &4.84 &-1.22 &4.9\\
AlP~(A3)	   &F$\overline{4}$3m	&5.463	&2.34	&-6.40 &2.62 &4.80 &2.51 &0.40 &2.5\\
AlAs~(A3)   &F$\overline{4}$3m	&5.661	&2.15	&-4.03 &2.40 &7.62 &2.29 &2.69 &2.23\\
AlSb~(A3)   &F$\overline{4}$3m	&6.136	&1.81	&6.50 &1.99 &17.75 &1.90 &12.42 &1.69\\
GaN~(A3)	   &F$\overline{4}$3m	&3.180	&3.17	&-3.35 &3.18 &-3.04 &3.13 &-4.57 &3.28\\
GaP~(A3)	   &F$\overline{4}$3m	&5.451	&2.29	&-2.55 &2.40 &2.12 &2.33 &-0.85 &2.35\\
%GaSb~(A3)   &F$\overline{4}$3m	&6.096	&0.91	&11.92 & & & & &0.82\\
GaAs~(A3)   &F$\overline{4}$3m	&5.648	&1.44	&-5.26 &1.88 &23.68 &1.83 &20.39 &1.52\\
InP~(A3)	   &F$\overline{4}$3m	&5.869	&1.52	&7.04 &1.82 &28.16 &1.77 &24.64 &1.42\\
InAs~(A3)   &F$\overline{4}$3m	&6.058	&0.53	&26.19 &0.92 &119.04 &0.88 &109.52 &0.42\\
InSb~(A3)   &F$\overline{4}$3m	&6.479	&0.53	&120.83 &0.99 &312.50 &0.96 &300.00 &0.24\\
%MgS~(A3)	   &F$\overline{4}$3m	&5.622	&4.78	&0.00 & & & & &4.78\\
MgTe~(A3)   &F$\overline{4}$3m	&6.420	&3.38	&-6.11 &3.80 &5.55 &3.70 &2.77 &3.6\\
CuCl~(A3)   &F$\overline{4}$3m	&5.501	&2.28	&-32.94 &2.37 &-30.29 &2.31 &-32.05 &3.4\\
CuBr~(A3)   &F$\overline{4}$3m	&5.820	&2.08	&-32.24 &2.31 &-24.75 &2.24 &-27.03 &3.07\\
CuI~(A3)	   &F$\overline{4}$3m	&6.063	&2.59	&-16.98 &2.91 &-6.73 &2.83 &-9.29 &3.12\\
ZnS~(A3)	   &F$\overline{4}$3m	&5.409	&3.32	&-13.54 &3.61 &-5.98 &3.52 &-8.33 &3.84\\
ZnSe~(A3)   &F$\overline{4}$3m	&5.668	&2.41	&-14.53 &2.82 &0.00 &2.74 &-2.83 &2.82\\
%ZnTe~(A3)   &F$\overline{4}$3m	&6.089	&	& & & & & &2.39\\
%MoS2   &F$\overline{4}$3m	&	&	& & & & & &1.29\\
AgI~(A3)	   &F$\overline{4}$3m	&6.499	&2.57	&-11.68 &2.84 &-2.40 &2.78 &-4.46 &2.91\\
CdS~(A3)	   &F$\overline{4}$3m	&5.818	&2.19	&-12.40 &2.45 &-2.40 &2.37 &-5.20 &2.5\\
CdSe~(A3)   &F$\overline{4}$3m	&6.052	&1.59	&-14.05 &1.96 &5.94 &1.89 &2.16 &1.85\\
CdTe~(A3)   &F$\overline{4}$3m	&6.480	&1.55	&-3.72 &2.01 &24.84 &1.94 &20.49 &1.61\\
%AlN~(A5)	   &P63mc				&a=3.111, c=4.978	&	& & & & & &6.19\\
%GaN~(A5)	   &P63mc				&a=3.180, c=5.166	&	& & & & & &3.5\\
%InN~(A5)	   &P63mc				&a=3.533, c=5.693	&	& & & & & &0.72\\
%BeO~(A5)	   &P63mc				&a=2.694, c=4.384	&	& & & & & &10.6\\
%ZnO~(A5)	   &P63mc				&a=3.258, c=5.220	&	& & & & & &3.44\\
%TiO$_2$~(A8)   &P42/mnm	        &a=4.594, c=2.959	&	& & & & & &3.3\\
%SrTiO$_3$~(A6) &Pm$\overline{3}$m	&3.901	&	& & & & & &3.3\\
%CuSCN  &R3m	                &	&	& & & & & &3.94\\
%SiO$_2$~(A9)   &P3121	            &a=4.921, c=5.400	&	& & & & & &9.65\\
\hline\hline
\end{tabular} 
\end{table*}

\begin{figure}[ht]
\begin{center}
\includegraphics[width=3.3in,height=3.0in,angle=0.0]{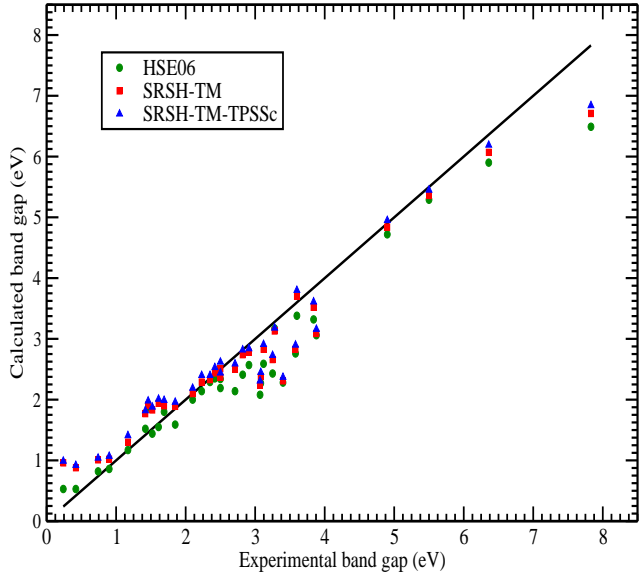} 
\end{center}
\caption{Calculated versus experimental band gaps for 36 solids presented in Table - II using different functionals.}
\label{fig2}
\end{figure}

\begin{table}%[h]
\label{latt-cons}
\caption{Summary statistics for the error in the calculated lattice constants and band gaps
for the set of solids presented in Table-I and Table-II.}
\begin{tabular}{c  c  c  c  c  c   c   c  c c c c}
\hline\hline
&HSE06&SRSH-TM-TPSSc&SRSH-TM\\
\hline
&&Lattice Constants\\
\hline 
%\hline
ME (\AA)&0.033&0.033&0.015\\
MAE (\AA)&0.039&0.052&0.042\\
%\hline
STDE(\AA)&0.069&0.079&0.075\\
MRE(\%)&0.582&0.536&0.207\\
MARE(\%)&0.731&0.959&0.815\\
STDRE(\%)&1.340&1.519&1.451\\
\hline
&&Band Gaps\\
\hline
ME (eV)&-0.306&-0.044&-0.121\\
MAE (eV)&0.371&0.329&0.331\\
%\hline
STDE (eV)&0.410&0.436&0.443\\
MRE(\%)&-4.144&14.917&11.202\\
MARE(\%)&15.878&24.934&23.392\\
STDRE(\%)&26.005&58.680&56.653\\
\hline\hline
\end{tabular} 
\end{table}

\section{Methodology}

The general scheme of inclusion of non-local XC potential  within the KS formalism is known as
generalized Kohn-Sham formalism (gKS). The gKS potential is written as, 

  \begin{equation}
 \begin{split}
    v_{xc}(\mathbf{r,r'}) = \alpha v_x^{HF-sr}(\mathbf{r,r'};\mu) +
    (1-\alpha)v_{x}^{sl-sr,\mu} \\+ v_{x}^{sl-lr,\mu} + v^{sl}_c~.
    \label{eq2}
    \end{split}
\end{equation}
Alternatively, this can be written as,
\begin{equation}
    v_{xc}(\mathbf{r,r'}) = \alpha v_x^{HF-sr}(\mathbf{r,r'};\mu) -\alpha v^{sl-sr,\mu}_x +
    v^{sl}_x + v^{sl}_c~,
    \label{eq3}
\end{equation}
 where the semilocal short range (sl-sr) and semilocal long range (sl-lr)
 part added into the semilocal exchange-correlation functional which in our present case is
 the TM functional. Here, the parameter $\alpha$ controls the amount of HF mixs with the
 semilocal functional and $\mu$ is the range separated parameter. The $\alpha=0$ value
 corresponds to pure semilocal formalism.

The range-separated density functional theory is actually developed by separating the
$\frac{1}{|\mathbf{r}-\mathbf{r}'|}$ operator into short and long-range part as,
 \begin{equation}
 \frac{1}{|\mathbf{r}-\mathbf{r}'|}=\underbrace{\frac{Erf(\mu
 |\mathbf{r}-\mathbf{r}'|)}{|\mathbf{r}-\mathbf{r}'|}}_{lr}
 +\underbrace{\frac{Erfc(\mu |\mathbf{r}-\mathbf{r}'|)}{|\mathbf{r}-\mathbf{r}'|}}_{sr}~.
 \label{eq4}
 \end{equation}
 Using the above seperation scheme the range separated parameter $\mu$ of Eq.(\ref{eq3}) is
 included into the exact HF exchange through the following equation,
\begin{equation}
 v_x^{HF-sr}(\mathbf{r,r'};\mu) =
 -\sum\limits_{i=1}^{\mathrm{{occ}}}\phi_i(\mathbf{r})\frac{Erfc(\mu|\mathbf{r}-\mathbf{r}'|)
 }{|\mathbf{r}-\mathbf{r}'|}\phi^*_i(\mathbf{r'}),
 \label{eq5}
\end{equation}
where $\phi_i$s  are single particle electronic orbitals. Except screened HF exchange, other
unknown potentials of Eq.(\ref{eq3}) are the screened potential ($v^{sl-sr,\mu}_x$) and the
semilocal potential ($v^{sl}_x$). In meta-GGA level theory the exchange potential is obtained
not only by taking the derivative with respect of density and gradient of density of the
exchange energy functional but also the partial derivative of KS kinetic energy density is
also required. In gKS formalism the semilocal exchange potential is expressed as 

\begin{eqnarray}
%\begin{split}
v^{sl}_{x}\Psi_{i} &=& \Big[\frac{\partial(\rho\epsilon_{x}^{sl})}{\partial\rho}
-\vec{\nabla}\frac{\partial(\rho\epsilon_{x}^{sl})}{\partial\vec{\nabla}\rho}\Big]
\Psi_{i}-\frac{1}{2}\vec{\nabla}\Big(\frac{\partial(\rho\epsilon_{x}^{sl})}{\partial\tau}\Big)
\vec{\nabla}\Psi_{i}\nonumber\\
&-&\frac{1}{2}\frac{\partial(\rho\epsilon_{x}^{sl})}{\partial\tau}\vec\nabla^2\Psi_{i}~,
\label{eq6}
%\end{split}
\end{eqnarray}
where $\epsilon_{x}^{sl}$ is the exchange energy density. In our 
present study $\epsilon_{x}^{sl}$ is the TM semilocal exchange 
energy density. The TM exchange energy functional can be expressed as,
\begin{eqnarray}
E_{x}^{TM}&=&-\int~d\mathbf{r}~\rho(\mathbf{r})\epsilon_{x}^{unif}F_{x}^{TM}~,\\
&=&-\int~d\mathbf{r}~\rho(\mathbf{r})\epsilon_{x}^{sl}~.
\label{eq2}
\end{eqnarray}
The TM enhancement factor is given by,
\begin{equation}
F_{x}^{TM}=wF_x^{DME} + (1-w)F_x^{sc}~,
\label{eq7}
\end{equation}
where, $F_x^{DME}=1/f^2+7R/(9f^4)$ is the enhancement factor derived from density matrix
expansion. Here, $R=1+595(2\lambda -1)^2p/54 -[\tau
-(3\lambda^2-\lambda+1/2)(\tau-\tau^{unif}-|\nabla
\rho|^2/(72\rho))]/\tau^{unif}$), $f = [1+10(70y/27) + \beta
y^2]^{1/10}$ (with $y=(2\lambda-1)^2p$) and the slowly varying fourth order gradient expansion
is given by 
$F_x^{sc}=\Big[1+10\Big\{\Big(\frac{10}{81}+\frac{50p}{729}\Big)p+
 \frac{146}{2025}\tilde{q}^2-\Big(\frac{73\tilde{q}}{405}\Big)\Big
 [\frac{3\tau^w}{5\tau}\Big](1-\frac{\tau^w}{\tau}\Big)\Big\}
 \Big]^{\frac{1}{10}}$. 
 The TM functional used $w$ as the weight factor between DME based functional form and slowly
 varying fourth order gradient expansion. The readers are suggested to go through the
 references~\cite{Tao-Mo16,YMo16} for the details of the derivation of the functional form and
 the terms associated with the TM functional. The interpolation factor $w$ is the function of
 meta-GGA  ingredient $z=\tau_W/\tau$, where $\tau_W$ is the von Weizs\"{a}cker kinetic energy
 density. In the slowly varying density limit $w$ is small therefore the fourth order density
 gradient approximation dominates. Not only that the interpolation factor $w$ have different
 unique features which makes the TM functionals works equally well both for molecular and
 solid state systems. The semilocal potential of the TM functional can be derived from
 Eq.(\ref{eq6}). Now, only the remaining part of the  potential is the  semilocal short-range
 part.  In the present case we have constructed the semilocal short-range from the LDA
 exchange hole. Using the LDA exchange hole the semilocal short range part of the  exchange
 energy functional becomes,
 \begin{equation}
 \begin{split}
 E_x^{sl-sr,\mu}=-\int~d\mathbf{r}~\rho(\mathbf{r})\epsilon_{x}^{unif}\Big\{1-\frac{8}{3}
 \mathcal{A}\Big(\sqrt{\pi}~erf(\frac{1}{2\mathcal{A}})\\+(2\mathcal{A}-4\mathcal{A}^3)
 e^{-\frac{1}
{4\mathcal{A}^2}}-3\mathcal{A} + 4\mathcal{A}^3\Big)\Big\},
\label{eq10}
\end{split}
\end{equation}
where $\epsilon_x^{unif} = \frac{3k_f}{4\pi}$ is the exchange energy per electron of the
homogeneous electron gas and $\mathcal{A} = \frac{\mu}{2k_f}$. Through $\mathcal{A}$ the
screening parameter includes into the semilocal short range part. The reason we use 
only LDA exchange hole in our semilocal short range part because, using it we obtain
satisfactory results for both the lattice constant and band gaps. Other way the inclusion of
exchange hole is given in reference~\cite{sjanapccp}. But the inclusion of full scheme (given
in reference~\cite{sjanapccp}) into this present functional form worsen its performance for
lattice constants. Therefore, we stick with the LDA exchange hole only. The semilocal short
range potential of the present functional form can also be obtained from Eq.(\ref{eq6}). This
completes the semilocal exchange potential of the present range separated functional. For the
correlation we have used one electron self-interaction free Tao-Perdew-Staroverov-Scuseria
(TPSS)~\cite{TPSS03} and its modified version~\cite{Tao-Mo16} for TM exchange in our present
study. The TPSS and TM correlation gives different results for various properties of
solids~\cite{YMo16}. Utilizing the present range separated functional coupled with the TPSS
and TM correlation therefore give rise two screened range separated functional.
We named those as (i)  SRSH-TM-TPSS  (screened range separated hybrid which  uses  TM 
exchange  plus TPSS correlation) and (ii) SRSH-TM (screened range separated hybrid which uses
TM exchange plus modified TPSS correlation). We assess the performance of both the
SRSH-TM-TPSS and SRSH-TM for solid state lattice constants and band gaps. 

All the self consistence calculations of the present functional is carried in the
projector-augmented-wave (PAW) environment with the plane wave basis set in Vienna Ab initio
simulation package (VASP)~\cite{paw1,paw2,vasp1,vasp2,vasp3,uspp}. The PAW methods are very 
accurate in density functional calculations and its performance is same as all electron
calculations used in different codes. The TM functional is recently implemented~\cite{jana} 
in VASP. Present implementation of range separate hybrid functional is based on this TM
implementation.  

Now we come to the discussion of the value of $\mu$ and $\alpha$ for our present range
separated functional. The value of the parameter $\alpha$ is chosen to be $0.25$. This
value is recommended by recently proposed TPSS based meta-GGA hybrid functional and also it
has been used in HSE06 functional. Regarding $\mu$ parameter, in the HSE06 functional the
$\mu$ parameter is set to $0.11$ bohr$^{-1}$, yielding a well balanced description for
lattice constants and band gaps. In the present case we also recommended $\mu=0.11$
bohr$^{-1}$ which produce a very balanced treatment of both the lattice constants and band
gaps. The performance of the present meta-GGA level range separated functional is compared
with widely used HSE06 functional. Unless otherwise stated the default values $\mu$ and
$\alpha$ values are used in VASP recommended HSE06 calculations. 

Here we have calculated the mean (relative) error (ME/MRE), mean absolute (relative) error
(MAE/MARE)  and the standard deviation of the (relative) error (STDE/STDRE) to study the
accuracy of each functionals. The definition we used here to calculate those are as follows,
\begin{eqnarray}
ME &=&\frac{1}{N}\sum_{i=1}^N(Y_i-y_i)\\
MAE &=& \frac{1}{N}\sum_{i=1}^N|Y_i-y_i|\\
MARE&=&\sum_i^N|Y_i-y_i|/|y_i|\\
STDRE&=&\Big[\sum_i^N(Y_i/y_i)-\frac{1}{N}(Y_i/y_i)\Big]^2,
\end{eqnarray}
where $Y_i$ and $y_i$ are the calculated and experimental values respectively.

%\section{Method of Calculation}
\section{Results and Discussions}
\subsection{Lattice Constants}

%---------------HSE06 DRAWBACK--------------------------

%Calculations of the lattice constants show that the fractional
%inclusion of exact exchange provides an improved
%description compared to the underlying semi-local PBE
%functional.8, 9 However, HSE06 inherits the PBE tendency to
%overestimate lattice constants, as well as the increase of the
%error for heavier elements. This overestimation can be pronounced
%for metals (e.g., Ag error 2%), ionic compounds
%(LiF, LiCl, NaF, NaCl approx. 1%), and heavier elements (α-
%Sn error > 1.2%).10 In addition, for transition metals, the atomization
%energies of solids exhibit significantly increased errors
%compared to PBE.

%-------------------------------------------------------------------

The fundamental test one should perform to check the robustness of a given functional for
solids is the equilibrium lattice constant. Predicting the accurate equilibrium lattice
constant is paramount important in view of  the structural properties of a solid. To perform
the benchmark calculation of SRSH-TM-TPSSc and SRSH-TM we employ the  two functionals for
studying 47 crystalline structures which includes a test set of six metals like Ag, Al, Cu,
Pd, K, Li and 41 semiconductors. Among the semiconductors we consider (i) 3 diamond
structures $-$ C, Si and Ge, (ii) 23 zinc blende structures $-$ SiC, BN, BP, BAs, BSb, AlP,
AlAs, AlSb, $\beta-$GaN, GaP, GaAs, GaSb, InP, InAs, InSb, ZnS, ZnSe, ZnTe, CdS, CdSe, CdTe,
MgS, MgTe, (iii) 17 ionic crystals $-$ MgO, MgSe, CaS, CaSe, CaTe, SrS, SrSe, SrTe, BaS,
BaSe, BaTe, LiCl, LiF, NaCl, LiF, NaCl, NaF. To test the performance and robustness of the
SRSH-TM-TPSSc and SRSH-TM we also put HSE06 into comparison. All the hybrid functional
calculations are performed starting from the well converged wavefunction of PBE calculation.
The $\Gamma-$ centered Monkhorst-Pack~\cite{mpsk} like 11$\times$11$\times$11 {\bf k} grids
are used for all our calculations.

In Table-I, we have summarized the performance of all the function under study. First we
discuss the performance of HSE06 functional. The HSE06 functional is based on the semilocal
PBE functional. Mixing fraction of exact exchange seems to be improve the lattice constant
compared to its semilocal form as it is shown in reference~\cite{jmhk}. It is well known that
HSE06 has the inherit tendency to overestimate lattice constants and it can be overcome using
a improved description of PBE i.e, PBEsol and its hybrid version HSEsol~\cite{shk}.
In TABLE III we have listed the overall statistics of HSE06 for all the solids using HSE06
functional. Overall, using HSE06 we obtain the MAE of 0.039\AA.

Now we come to the performance of newly constructed SRSH-TM-TPSSc and SRSH-TM. Regarding the 
performance of SRSH-TM-TPSSc, due to the TPSS correlation the lattice constants of all the 
crystalline structures are overestimated in this case. This drawback can be explained from
the performance of TM-TPSS as reported in reference~\cite{jana}. In the reference~\cite{jana}
it is shown that MAE of TM-TPSS is 0.045\AA. Mixing HF with TM-TPSS actually overestimates
more and give the MAE of 0.045\AA. In the present formalism of screened hybrid functional
theory using TM correlation with HF improves its over SRSH-TM-TPSSc and gives MAE 0.042\AA~
which is only 0.004\AA~ and 0.003\AA~ greater than its base functional TM and hybrid HSE06
respectively. Regarding the performance of HSE06 and semilocal TM functional both perform
equivalently as it is shown in this paper and reference~\cite{jana}. From Fig.(\ref{fig1}) it
is evident that in case of Ge, AlAs, GaP, GaAs, InP, InAs, InSb, ZnS, ZnSe, ZnTe, CdTe, MgS,
MgSe, MgTe, Ag, Cu, K the SRSH-TM actually performs better than HSE06. For other cases the
performance of HSE06 is better or equivalent compared to SRSH-TM.

%One can also compare present SRSH-TM-TPSSc and SRSH-TM with its base
%functional i.e., TM-TPSS and TM. Using TM-TPSS and TM functional we obtain MAE 0.045 \AA~and 0.38 \AA~respectively which are very accurate within the semilocal formalism.  (Next discuss effect of inclusion of exact exchange with TM-TPSS and TM) 

%Now we will discuss the performance of HSE06 with meta-GGA level screened hybrid SRSH-TM-TPSSc and SRSH-TM. (In this paragraph discuss the comparison of SRSH-TM-TPSSc and SRSH-TM with HSE06.)

\subsection{Band Gaps}

%\subsection{The DME-sc-TPSSc/TMc functional}

It is well known that the accurate band gap prediction is only achievable through the hybrid
functional scheme due to the inclusion of HF exchange which actually balance the
delocalization and localization problem arises from semilocal and HF exchange. As present
range separated scheme is based on the hybrid interface, therefore, it is always interesting
to check the performance of the present scheme for band gaps problem. Besides the present
scheme is based on the semilocal funcional which is the second best performer after SCAN
meta-GGA in predicting the band gaps within semilocal formalism~\cite{jana}. Also, it is
noteworthy to mentioned that the meta-GGA functional implemented with the framework of gKS
formalism gives more realistic band gap~\cite{relbg}. Therefore, within meta-GGA hybrids the
improvement in band gap comes from both the semilocal formalism and mixing of HF exchange.

Here, we choose 34 semiconductors (including insulators) to assess the performance of
SRSH-TM-TPSSc and SRSH-TM along with the HSE06. We report performance of all the functionals
in TABLE II. Here, all the band gaps are calculated at experimental lattice constants. The
experimental geometries are collected from reference~\cite{tranbg}. From 
TABLE II it is evident that all the screened hybrid functionals perform better than their
respective semilocal form due to the inclusion of HF exchange. It is well known that, the
HSE06 functional is widely used functional for prediction band gap of semiconductor for small
band gap material (upto 5 eV). For large band gap materials HSE06 actually underestimates the
band gap. But, the performance of HSE06 quite productive because the computational cose of
HSE06 is less than accurate many body treatment like GW or many body perturbation theory
(MBPT). Regarding the performance of meta-GGA level screened range separated hybrids
SRSH-TM-TPSSc and SRSH-TM, we observed the band gap is more enhanced
than HSE06 for all the materials. this is obvious because the meta-GGA functionals are 
implemented in gKS scheme which produced more realistic band gap than GGA. Here, mixing the
HF exchange with meta-GGA level semilocal functionals actually enhance the band gap more. For
the semiconductors for which HSE06 underestimates the band gap slightly, inclusion of HF
exchange with meta-GGA level screened range separation scheme actually compensate those.
Regarding the overall comparison of SRSH-TM-TPSSc with SRSH-TM and HSE06, the SRSH-TM-TPSSc
overestimate the band gaps of those materials for which HSE06 and SRSH-TM quite accurate. It
has been observed for the band gap values of ScN, Si, Ge, SiC, GaP, InP, InAs, InSb, and 
CdTe that HSE06 is accurate for those systems. SRSH-TM also gives very comparable results
with HSE06. But, the overestimation of band gaps is observed using SRSH-TM-TPSSc for those
materials. Overall, the SRSH-TM functional is quite productive over HSE06 for predicting the
band gap of semiconductor materials. It is also noteworthy to mention that, the calculated
band gap values at their experimental lattice constant can also be compared with different
semilocal functionals which are quite good (but not always) in predicting band gaps in
semilocal level~\cite{tranbg}. Interestingly, few band gaps reported using TPSS based
screened hybrid functionals in reference~\cite{tpssrs} is also comparable with the band gap
of SRSH-TM in TABLE II. 

The only drawback of present functional form is that it overestimates the band gap for which
HSE06 is exact (or slightly overestimating). It could be avoided by using the full reverse
engineered exchange hole of TM functional. It also noteworthy that TPSS based screened
functional also overestimates those values as shown for few specific cases in
reference~\cite{tpssrs}.

%\subsection{Lattice Constants}

\section{Conclusions and future direction}

We assess the performance of solid state lattice constants and band gaps using the meta-GGA
level screened range separated hybrids SRSH-TM-TPSSc and SRSH-TM with the PAW method. To
check the robustness of the present proposition we also compared the performance of the
present functional with HSE06 functional. The results obtained using the SRSH-TM-TPSSc and
SRSH-TM are found to be interesting. This is first ever test of any meta-GGA level screened 
range separated hybrid functional for both the lattice constants and band gaps. From the
prospective of lattice constants the overall performance of SRSH-TM is quite impressive.
The MAE of SRSH-TM differs from HSE06 and TM semilocal functional only by 0.003\AA~ and
0.004\AA. It has been observed that for several cases where HSE06 has the tendency to
overestimate the lattice constants SRSH-TM performs quite well. Regarding the performance of
SRSH-TM and SRSH-TM-TPSSc, the performance of SRSH-TM is quite better compare to
SRSH-TM-TPSSc. The improved performance of SRSH-TM actually comes from change is correlation.
TM exchange coupled with TM correlation performs better than TPSS correlation for solid state
lattice constants.

From the point of view of functional form it is very simple. Only the LDA based exchange hole
is used in its short range together with HF short range exchange. It is observed that a
simple modification on the top of the TM functional improves its performance for band gap.
As discussed earlier the band gap problem is not achievable through the semilocal level only
suitably mixing HF with semilocal performs well for band gaps. The performer of SRSH-TM
indicates that it is a good competitor with HSE06, especially for cases HSE06 has the
tendency to underestimate the band gaps. Though few cases SRSH-TM overestimates the band gap
more compare to HSE06. But the overall performance of SRSH-TM is quite well.

Lastly, we want to conclude that the present SRSH functional based on TM semilocal functional
keeps all the good properties of TM functionals which is very accurate for predicting solid
state properties in semilocal level. The advantage of the present SRSH-TM (we recommand
SRSH-TM over SRSH-TM-TPSSc because SRSH-TM performances more balanced way than
SRSH-TM-TPSSc for both the lattice constant and band gap) is that is mixes HF which actually
overcome several drawback that is actually not achievable in semilocal level for TM, more
precisely the band gap. Another interesting feature of the present formalism is that it is
based on meta-GGA level theory which is very accurate than GGA in different ways. Several 
electronic structural properties can be studied using this screened meta-GGA level
functional. As a future direction of present SRSH functional it is always interesting to
study the dielectric dependent performance of SRSH because it improves the screening effects
and several other properties~\cite{galli1,galli2,galli3,galli4}.

\end{document}